# A note on tree factorization and no particle production

Klaus Bering[a]

Institute for Theoretical Physics & Astrophysics
Masaryk University
Kotlářská 2
CZ–611 37 Brno
Czech Republic

July 26, 2022

**Abstract**

We prove factorization of the generating functional of connected tree diagrams by exploring that it is the Legendre transform of the action. This theorem is then applied to the example of a massive real scalar field theory in 2D. In the process we streamline the proof that the assumption of no particle production leads to either the sin(h)-Gordon or the Bullough-Dodd model.

---

[a]E–mail: `bering@physics.muni.cz`

# Contents

# 1 Introduction

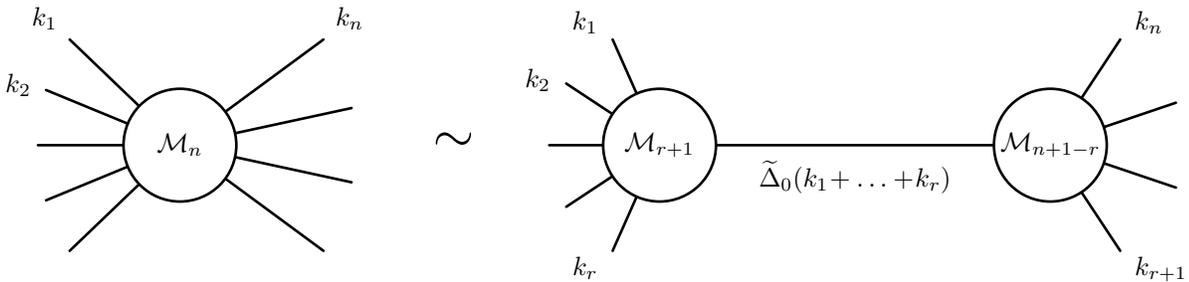

Figure 1: Factorization of connected tree amplitude $\mathcal{M}_n$ for $(k_1 + \ldots + k_r)^2 + m^2 \to 0$, cf. theorem 2.8. Outgoing wavevectors (=momenta/$\hbar$) of the external legs are denoted by $k_1, k_2, \ldots, k_n$.

This paper consists of 2 parts. In the first part (section 2), we prove the factorization of a connected tree amplitude into two subamplitudes, cf. Fig. 1. Here it is a non-trivial fact that the 2 subamplitudes are in fact amplitudes in their own rights, and not, say, proper subsets of Feynman diagrams of those. To systematically manage the diagrammatical combinatorics, we explore in a novel way the Legendre transformation between the action and the generating functional of connected tree diagrams.

The paper will focus on tree diagrams, which are the leading diagrams in a semi-classical $\hbar$-expansion. For a modern review of factorization at loop-level and its relation to (generalized) unitarity, see e.g. Ref. [1].

In the second part (section 3), we apply tree factorization to a scalar field theory in 2D, and study when the tree amplitudes has no particle production, i.e. that the connected $n$-point tree amplitude $\mathcal{M}_n = 0$ vanishes for $n \geq 5$. The latter approach dates back to the seminal works of Refs. [2, 3]. For more recent works, see e.g. Refs. [4, 5, 6]. By systematically considering conditions on $\mathcal{M}_n$ order-by-order in $n$, we give a streamlined complete proof that there are only 2 such theories:

1. The sin(h)-Gordon model.

2. The Bullough-Dodd model [7].

## 2 Connected trees and factorization

### 2.1 Generating functional $W_c^{\text{tree}}[J]$ of connected trees

Let a theory have an action $S[\phi]$.

**Theorem 2.1** *If the action $S[\phi]$ does not depend on Planck's constant $\hbar$, and if the Euler-Lagrange (EL) equations*[1]

$$J_k \approx -\frac{\delta S[\phi]}{\delta \phi^k} \qquad \Leftrightarrow \qquad \phi^k \approx \phi^k[J] \tag{2.1}$$

*has a unique solution for given sources $J_k$ (and boundary conditions), then the Legendre transform of $S[\phi]$ is the generating functional*

$$W_c^{\text{tree}}[J] = S[\phi] + J_k \phi^k \tag{2.2}$$

*of connected tree diagrams. In particular*

$$J_k = -\frac{\delta S[\phi]}{\delta \phi^k} \qquad \Leftrightarrow \qquad \phi^k = \frac{\delta W_c^{\text{tree}}[J]}{\delta J_k}. \tag{2.3}$$

PROOF OF THEOREM 2.1:

$$\begin{aligned} \exp\left\{\frac{i}{\hbar} W_c[J]\right\} &= Z[J] = \int \mathcal{D}\frac{\phi}{\sqrt{\hbar}} \exp\left\{\frac{i}{\hbar}\left(S[\phi] + J_k \phi^k\right)\right\} \\ &\sim \text{Det}\left(\frac{1}{i}\frac{\delta^2 S[\phi[J]]}{\delta \phi^m \delta \phi^n}\right)^{-1/2} \exp\left\{\frac{i}{\hbar}\left(S[\phi[J]] + J_k \phi^k[J]\right)\right\}(1 + \mathcal{O}(\hbar)) \end{aligned} \tag{2.4}$$

In the first equality, we used that the exponential of the generating functional $W_c[J]$ of connected diagrams yields all diagrams, i.e. the path integral $Z[J]$. At the last step we used the stationary phase/WKB approximation $\hbar \to 0$. Finally, use the $\hbar$/loop-expansion to deduce the theorem: The dominating contributions on the 2 sides of eq. (2.4) is precisely eq. (2.2) □

---

[1] We use DeWitt's condensed notation, i.e. the index $k$ is a collection of discrete and continuous indices, and repeated indices are summed/integrated over.

## 2.2 Perturbative expansions

Let us expand perturbatively the action

$$S[\phi] = \sum_{n=0}^{\infty} S_n[\phi], \qquad S_n[\phi] = \frac{1}{n!}(S_n)_{k_1,\ldots k_n}\phi^{k_1}\ldots\phi^{k_n}; \qquad (2.5)$$

the generating functional of connected trees

$$W_c^{\text{tree}}[J] = \sum_{n=0}^{\infty} W_{c,n}^{\text{tree}}[J], \qquad W_{c,n}^{\text{tree}}[J] = \frac{1}{n!}(W_{c,n}^{\text{tree}})^{k_1,\ldots k_n} J_{k_1}\ldots J_{k_n}; \qquad (2.6)$$

and the generating functional of rooted trees/classical solution

$$\phi^k[J] = \sum_{n=0}^{\infty} \phi_n^k[J], \qquad \phi_n^k[J] = \frac{1}{n!}(\phi_n^k)^{k_1,\ldots k_n} J_{k_1}\ldots J_{k_n}. \qquad (2.7)$$

**Assumption 2.2** *We will always assume that there are no classical tadpoles*

$$(S_1)_k = 0 \quad \Leftrightarrow \quad (W_{c,1}^{\text{tree}})^k = 0 \quad \Leftrightarrow \quad \phi_0^k = 0. \qquad (2.8)$$

Then there are no connected vacuum trees

$$W_{c,0}^{\text{tree}} = S_0, \qquad (2.9)$$

apart from the action constant $S_0$; and the quadratic terms are given by the free propagator

$$S_2[\phi] = -\frac{1}{2}\phi^k(\Delta_0^{-1})_{k\ell}\phi^\ell \quad \Leftrightarrow \quad W_{c,2}^{\text{tree}}[J] = \frac{1}{2}J_k(\Delta_0)^{k\ell}J_\ell \quad \Leftrightarrow \quad \phi_1^k[J] = (\Delta_0)^{k\ell}J_\ell. \qquad (2.10)$$

It is straightforward to generate formulas for the tree diagrams in terms of a finite number of building blocks consisting of free propagators $(\Delta_0)^{k\ell}$ and $n$-vertices $(S_n)_{k_1,\ldots k_n}$ to each order in perturbation theory.

## 2.3 Vertex expansion of a rooted tree

**Theorem 2.3**

$$\begin{aligned}
\frac{\delta W_{c,\geq 3}^{\text{tree}}[J]}{\delta J_k} &\stackrel{(2.3)}{=} \phi_{\geq 2}^k[J] \\
&= \phi^k[J] - (\Delta_0)^{k\ell}J_\ell \\
&\stackrel{(2.3)}{=} (\Delta_0)^{k\ell}\left(-\frac{\delta S_2[\phi]}{\delta \phi^\ell} + \frac{\delta S[\phi]}{\delta \phi^\ell}\right)\bigg|_{\phi=\phi[J]} \\
&= (\Delta_0)^{k\ell} \frac{\delta S_{\geq 3}[\phi]}{\delta \phi^\ell}\bigg|_{\phi = \frac{\delta W_c^{\text{tree}}[J]}{\delta J}} \\
&= (\Delta_0)^{k\ell} \frac{\delta S_{\geq 3}[\phi]}{\delta \phi^\ell}\bigg|_{\phi = \Delta_0 J + \frac{\delta W_{c,\geq 3}^{\text{tree}}[J]}{\delta J}}.
\end{aligned} \qquad (2.11)$$

**Remark 2.4** *Theorem 2.3 can be interpreted diagramatically as a recursion relation: The root-leg of the rooted tree $\delta W_{c,\geq 3}^{\text{tree}}[J]/\delta J_k$ is connected via a free propagator $\Delta_0$ to a vertex in the action $S_{\geq 3}$. The other lines of this vertex are either i) an external leg or ii) an internal line that is the beginning of a new rooted tree $\delta W_{c,\geq 3}^{\text{tree}}[J]/\delta J_k$. One should sum over possible types of vertices in the action $S_{\geq 3}$. For an example, see Fig. 13.*

## 2.4 Tree factorization

Let us consider a connected tree $\Gamma$ with 2 types of sources $J$ and $K$. For each line/propagator $\Delta_0$ in $\Gamma$ the tree $\Gamma = \Gamma'\Delta_0\Gamma''$ divides in two subtrees $\Gamma'$ and $\Gamma''$.

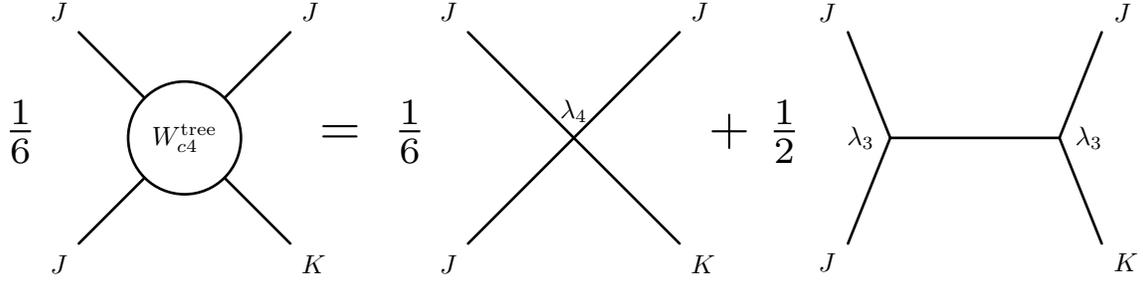

Figure 2: The left–hand side is an example of a mixed connected 4-point tree correlator function: Each diagram is divided by its symmetry factor. The lines attached to the $K$ source on the right–hand side are factorizable.

**Definition 2.5** *If $\Gamma'$ does not depend on $J$ and $\Gamma''$ does not depend on $K$ (or vice-versa), then the line/propagator $\Delta_0$ is called* **factorizable**. *If a tree $\Gamma$ has no factorizable lines, it is called* **non-factorizable**.

We can now formulate the main theorem 2.6.

**Theorem 2.6**

$$W^{\text{tree}}_{c,\geq 2}[J+K] - W^{\text{tree}}_{c,\geq 2}[J] - W^{\text{tree}}_{c,\geq 2}[K] \;=\; \frac{\delta W^{\text{tree}}_c[J]}{\delta J_k}(\Delta_0^{-1})_{k\ell}\frac{\delta W^{\text{tree}}_c[K]}{\delta K_\ell} + \text{non-factorizable trees}. \qquad (2.12)$$

**Remark 2.7** *A tree with source $J + K$ is to be viewed as a sum of trees with two types of sources $J$ and $K$ via the binomial formula. An example of a term in such binomial expansion is illustrated in Fig. 2.*

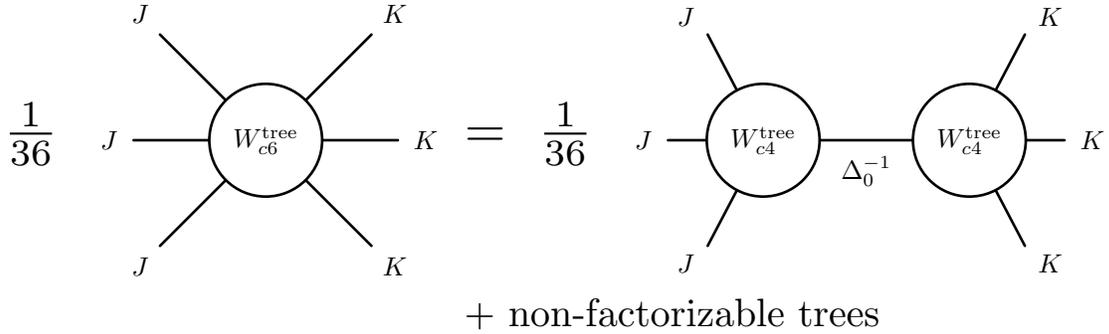

+ non-factorizable trees

Figure 3: An example of factorization of a mixed connected 6-point tree correlator function: Each diagram is divided by its symmetry factor.

PROOF OF THEOREM 2.6: Define mixed terms

$$\phi^k_{\geq 2}[J,K] \;:=\; \phi^k_{\geq 2}[J+K] - \phi^k_{\geq 2}[J] - \phi^k_{\geq 2}[K]. \qquad (2.13)$$

Diagrammatically eq. (2.13) consists of rooted trees with both $J$ and $K$ leaves. Then we may expand

$$\begin{aligned}
\phi^k[J+K] &= \phi_1^k[J+K] + \phi_{\geq 2}^k[J+K] \\
&= \phi_1^k[J] + \phi_1^k[K] + \phi_{\geq 2}^k[J+K] \\
&\stackrel{(2.13)}{=} \phi^k[J] + \phi^k[K] + \phi_{\geq 2}^k[J,K].
\end{aligned} \quad (2.14)$$

Hence

$$\begin{aligned}
W_{c,\geq 2}^{\text{tree}}[J+K] &\stackrel{(2.2)}{=} S_{\geq 2}[\phi[J+K]] + (J_k + K_k)\phi^k[J+K] \\
&\stackrel{(2.14)}{=} S_{\geq 2}\left[\phi[J] + \phi[K] + \phi_{\geq 2}[J,K]\right] + (J_k + K_k)\phi^k[J+K] \\
&\stackrel{(2.14)}{=} S_{\geq 2}\left[\phi[J] + \phi[K]\right] + \left.\frac{S_{\geq 2}[\phi]}{\delta\phi^k}\right|_{\phi=\phi[J]+\phi[K]} \phi_{\geq 2}^k[J,K] \\
&\quad + \mathcal{O}\left(\phi_{\geq 2}[J,K]^2\right) + (J_k + K_k)\left(\phi^k[J] + \phi^k[K] + \phi_{\geq 2}^k[J,K]\right) \\
&= S_{\geq 2}[\phi[J]] + S_{\geq 2}[\phi[K]] + \frac{S_{\geq 2}[\phi[J]]}{\delta\phi^k}\phi^k[K] + \phi^k[J]\frac{S_{\geq 2}[\phi[K]]}{\delta\phi^k} \\
&\quad + \phi^k[J](\Delta_0^{-1})_{k\ell}\phi^\ell[K] + \mathcal{O}\left(\phi[J]^2\phi[K]^2\right) \\
&\quad + \left(\frac{S[\phi[J]]}{\delta\phi^k} + \frac{S[\phi[K]]}{\delta\phi^k} + \mathcal{O}(\phi[J]\phi[K]) + J_k + K_k\right)\phi_{\geq 2}^k[J,K] \\
&\quad + (J_k + K_k)\left(\phi^k[J] + \phi^k[K]\right) \\
&\stackrel{(2.3)}{=} S_{\geq 2}[\phi[J]] + S_{\geq 2}[\phi[K]] + \phi^k[J](\Delta_0^{-1})_{k\ell}\phi^\ell[K] \\
&\quad + J_k\phi^k[J] + K_k\phi^k[K] + \text{non-factorizable trees} \\
&\stackrel{(2.2)}{=} W_{c,\geq 2}^{\text{tree}}[J] + W_{c,\geq 2}^{\text{tree}}[K] + \frac{\delta W_c^{\text{tree}}[J]}{\delta J_k}(\Delta_0^{-1})_{k\ell}\frac{\delta W_c^{\text{tree}}[K]}{\delta K_\ell} \\
&\quad + \text{non-factorizable trees.}
\end{aligned} \quad (2.15)$$

Here we have used that if either

$$\mathcal{O}\left(\phi_{\geq 2}[J,K]^2\right), \qquad \mathcal{O}\left(\phi[J]\phi[K]\phi_{\geq 2}[J,K]\right), \qquad \mathcal{O}\left(\phi[J]^2\phi[K]^2\right), \quad (2.16)$$

are attached to a vertex then the tree diagram must be non-factorizable. $\square$

## 2.5 On-shell tree-level scattering amplitude

Amputated/stripped amplitudes $\mathcal{M}_n(k_1,\ldots,k_n)$ of connected trees are obtained from the generating functional $W_c^{\text{tree}}[J]$ as follows:[2]

$$\frac{i}{\hbar}\mathcal{M}_n(k_1,\ldots,k_n)\,(2\pi)^d\delta^d(k_1+\ldots+k_n) = \widetilde{\Delta}_0^{-1}(k_1)\frac{\delta}{\delta\widetilde{J}(k_1)}\ldots\widetilde{\Delta}_0^{-1}(k_n)\frac{\delta}{\delta\widetilde{J}(k_n)}W_c^{\text{tree}}[J]\bigg|_{J=0}, \quad (2.17)$$

where $n \geq 3$. Let us analytically continue $\mathcal{M}_n(k_1,\ldots,k_n)$ for complex outgoing wavevectors $k_i$ (=momenta/$\hbar$). The $n$-point tree-amplitude $\mathcal{M}_n$ is a rational function of $k_i$. It will become crucial in what follows that poles in $\mathcal{M}_n$ can only come from poles in internal propagators.

On-shell condition:[3]

$$k_i^2 + m_i^2 \approx 0. \quad (2.18)$$

Theorem 2.6 now transcribes into the following tree factorization theorem 2.8, cf. Fig. 1.

---

[2]The tildes denote Fourier transform.
[3]We use the Minkowski signature $(-,+,\ldots,+)$. The mass parameters $m_i$ are assumed to have dimension of inverse length. The $\approx$ symbol denotes in principle an on-shell equality, although we will not always use the notation.

**Theorem 2.8** *Tree factorization:*

$$\mathcal{M}_n(k_1,\ldots,k_n) \sim \mathcal{M}_{r+1}(k_1,\ldots,k_r,\cdot)\,\widetilde{\Delta}_0(k_1+\ldots+k_r)\,\mathcal{M}_{n+1-r}(k_{r+1},\ldots,k_n,\cdot) + \text{finite terms}$$
$$\text{for} \quad (k_1+\ldots+k_r)^2 + m^2 \to 0 \tag{2.19}$$

*if the factorization channel is kinematically unique.*

**Remark 2.9** *Be aware that on one hand a mixed $W_c^{\text{tree}}$ correlator function has always a unique JK-factorization, while on the other hand the limit $(k_1+\ldots+k_r)^2 + m^2 \to 0$ in the on-shell tree amplitude $\mathcal{M}_n(k_1,\ldots,k_n)$ may correspond to several[4] factorization channels simultaneously if the number of external legs is low. Hence the latter may not be factorizable, cf. counterexamples in subsections 3.5.1 and 3.6.2.*

## 3 Example: No particle production in 2D

We now apply the tree factorization theorems of section 2 to a scalar field theory in 2D with no particle production.

### 3.1 Scalar field theory in 2D

The Lagrangian density for a relativistic massive real scalar field in $d=2$ is

$$\mathcal{L} = -\frac{1}{2}\partial_\mu\phi\partial^\mu\phi - \mathcal{V} = \partial_+\phi\,\partial_-\phi - \mathcal{V}, \qquad \mathcal{V} = \frac{1}{2}m^2\phi^2 + \sum_{n\geq 3}\frac{\lambda_n}{n!}\phi^n,$$
$$m > 0, \qquad \lambda_n = m^2 g_n, \qquad \widetilde{\Delta}_0(k) = \frac{1}{k^2 + m^2 - i\epsilon}. \tag{3.1}$$

The $i\epsilon$ prescription is implicitly implied from now on. We exclude interaction terms with spacetime derivatives. All coupling constants $\lambda_n = m^2 g_n$ have dimension of inverse quadratic length, i.e. are relevant/renormalizable.[5]

It will be convenient to use light-cone coordinates

$$ds^2 = -dt^2 + dx^2 = -2dx^+ dx^-, \qquad x^\pm = \frac{t \pm x}{\sqrt{2}},$$
$$(k_i^+, k_i^-) \approx \frac{m}{\sqrt{2}}(a_i, a_i^{-1}), \qquad a_i \in \mathbb{C}\backslash\{0\}. \tag{3.2}$$

The tree amplitudes $\mathcal{M}_n$ are Lorentz/boost invariant $a_i \to e^\eta a_i$ since they are built from Lorentz invariant propagators and vertices. It will become important in what follows that if they don't have poles, they must be constant.

**Remark 3.1** *No particle production here means that the outgoing particles are a permutation of the ingoing particles. This implies that $\mathcal{M}_{n\geq 5} = 0$ must vanish. E.g. the connected $3 \to 3$ tree-amplitude must vanish because it is an analytic continuation of, say, the connected $2 \to 4$ tree-amplitude.*

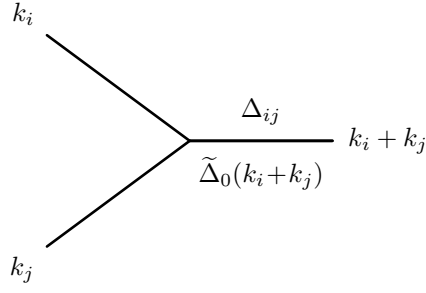

Figure 4: A 2-propagator $\widetilde{\Delta}_0(k_i+k_j) = -m^{-2}\Delta_{ij}$ is by definition attached to 2 external legs.

## 3.2 Propagators

### 3.2.1 2-propagator

The 2-propagator is

$$\begin{aligned}
-m^2\widetilde{\Delta}_0(k_i+k_j) &= \frac{-m^2}{(k_i+k_j)^2+m^2} \stackrel{(3.2)}{\approx} \frac{1}{(a_i+a_j)(a_i^{-1}+a_j^{-1})-1} \\
&= \frac{a_i a_j}{(a_i+a_j)^2 - a_i a_j} = \frac{a_i a_j}{a_j^2 + a_i a_j + a_j^2} \\
&= \frac{a_i a_j}{(a_i - e^{2\pi i/3}a_j)(a_i - e^{-2\pi i/3}a_j)} =: \Delta_{ij},
\end{aligned} \qquad (3.3)$$

cf. Fig. 4. Poles:

$$a_i = e^{\pm 2\pi i/3} a_j. \qquad (3.4)$$

### 3.2.2 3-propagator

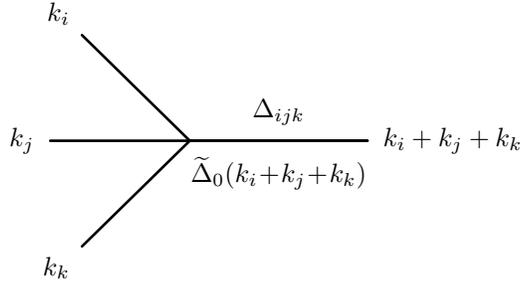

Figure 5: A 3-propagator $\widetilde{\Delta}_0(k_i+k_j+k_k) = -m^{-2}\Delta_{ijk}$ is by definition attached to 3 external legs.

---

[4]After diagonalizing the mass matrix, there is of course always a sum over degenerate masses.
[5]The only divergencies in diagrams come from self-loops, which are logarithmically divergent. These can be removed in the interaction picture of the operator formalism via normal ordering [8].

The 3-propagator is

$$-m^2 \widetilde{\Delta}_0(k_i+k_j+k_k) = \frac{-m^2}{(k_i+k_j+k_k)^2 + m^2}$$
$$\overset{(3.2)}{\approx} \frac{1}{(a_i+a_j+a_k)(a_i^{-1}+a_j^{-1}+a_k^{-1}) - 1}$$
$$= \frac{a_i a_j a_k}{(a_i+a_j)(a_j+a_k)(a_k+a_i)} =: \Delta_{ijk},$$
(3.5)

cf. Fig. 5. Poles:

$$a_i + a_j = 0. \tag{3.6}$$

Residue:

$$\mathrm{Res}(\Delta_{ijk}, a_i+a_j) = \frac{a_i^2 a_k}{a_i^2 - a_k^2}. \tag{3.7}$$

### 3.3   3-point amplitude

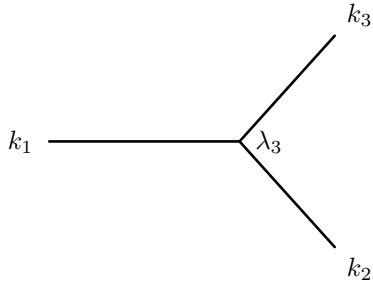

Figure 6: Connected 3-point tree diagrams.

The 3-point tree amplitude is[6]

$$\mathcal{M}_3(k_1, k_2, k_3) = -\lambda_3, \tag{3.8}$$

cf. Fig. 6. Momentum conservation implies complex (and hence unphysical) lightcone momenta

$$\left. \begin{array}{r} a_1 + a_2 + a_3 = 0 \\ a_1^{-1} + a_2^{-1} + a_3^{-1} = 0 \end{array} \right\} \quad \Rightarrow \quad (a_1, a_2, a_3) \propto (1, e^{\pm 2\pi i/3}, e^{\mp 2\pi i/3}). \tag{3.9}$$

$\mathcal{M}_3$ is constant but unphysical. In particular, the 3-point function $\mathcal{M}_3$ has no particle production despite being non-zero.

### 3.4   4-point amplitude

The 4-point tree amplitude is

$$\mathcal{M}_4(k_1, k_2, k_3, k_4) = -\lambda_4 + \lambda_3^2 \left( \widetilde{\Delta}_0(k_1+k_2) + \widetilde{\Delta}_0(k_2+k_3) + \widetilde{\Delta}_0(k_3+k_1) \right) \tag{3.10}$$

$$\Rightarrow \quad -m^{-2} \mathcal{M}_4(k_1, k_2, k_3, k_4) = g_4 + g_3^2 (\Delta_{12} + \Delta_{23} + \Delta_{31}), \tag{3.11}$$

cf. Fig. 7. Momentum conservation restricts the 4 lightcone-momenta $(a_1, a_2, a_3, a_4)$ to 2 independent parameters:

$$\left. \begin{array}{r} a_1 + a_2 + a_3 + a_4 = 0 \\ a_1^{-1} + a_2^{-1} + a_3^{-1} + a_4^{-1} = 0 \end{array} \right\} \quad \Rightarrow \quad (a_1, a_2, a_3, a_4) \text{ are 2 pairs of opposite values } a_i + a_j = 0. \tag{3.12}$$

---

[6]The minus in eq. (3.8) is because of a minus in eq. (3.1).

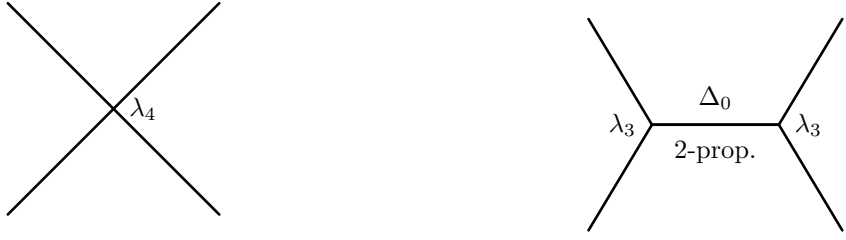

Figure 7: Connected 4-point tree diagrams.

Eq. (3.12) implies that the 4-point function $\mathcal{M}_4$ has no particle production.

If we e.g. assume that $a_1 + a_3 \approx 0$, then the *stu* channel simplifies to

$$\begin{aligned} a_1 + a_3 \approx 0 \quad \Rightarrow \quad \Delta_{12} + \Delta_{23} + \Delta_{31} &\approx \frac{a_1 a_2}{(a_1+a_2)^2 - a_1 a_2} - \frac{a_1 a_2}{(a_1-a_2)^2 + a_1 a_2} - 1 \\ &= \frac{a_1 a_2}{a_1^2 + a_1 a_2 + a_2^2} - \frac{a_1 a_2}{a_1^2 - a_1 a_2 + a_2^2} - 1 \\ &= -\frac{2 a_1^2 a_2^2}{(a_1^2 + a_2^2)^2 - a_1^2 a_2^2} - 1. \end{aligned} \quad (3.13)$$

$\mathcal{M}_4$ is non-constant and has unphysical poles at $a_1 = \pm' e^{\pm 2\pi i/3} a_2$ caused by a 2-propagator iff $g_3 \neq 0$. $\mathcal{M}_4$ factorizes $\mathcal{M}_4 \sim \mathcal{M}_3 \Delta_0 \mathcal{M}_3$ on these poles, cf. theorem 2.8.

### 3.5 5-point amplitude

#### 3.5.1 No single-channel $\mathcal{M}_5 \sim \mathcal{M}_4 \Delta_0 \mathcal{M}_3$ factorization

Consider e.g. a pole $a_4 = e^{\pm 2\pi i/3} a_5$ associated with a 2-propagator for the channel 45. Define

$$a_{45} := a_4 + a_5 = -e^{\mp 2\pi i/3} a_5. \quad (3.14)$$

Momentum conservation then implies

$$\Rightarrow \quad (a_1, a_2, a_3, a_{45}) \text{ are 2 pairs of opposite values } a_i + a_j = 0. \quad (3.15)$$

In other words: $\exists i \in \{1,2,3\} : \; a_i = e^{\mp 2\pi i/3} a_5$. This leads to a simultaneous pole for another channel $i5$.

So factorization is not useful for the calculation of $\mathcal{M}_5$. However, we don't have to actually calculate $\mathcal{M}_5$ itself: It is enough to calculate the residues of the poles [4].

#### 3.5.2 5-point amplitude

The 5-point tree amplitude is

$$\begin{aligned} \mathcal{M}_5(k_1, k_2, k_3, k_4, k_5) = &- \lambda_5 + \lambda_3 \lambda_4 \sum_{\substack{i,j,k \in \{1,\ldots,5\} \\ i<j<k}} \widetilde{\Delta}_0(k_i + k_j + k_k) \sum_{\substack{\ell,m \notin \{i,j,k\} \\ \ell<m}} 1 \\ &- \frac{\lambda_3^3}{2} \sum_{i=1}^{5} \sum_{\substack{j,k \notin \{i\} \\ j<k}} \sum_{\substack{\ell,m \notin \{i,j,k\} \\ \ell<m}} \left\{ \begin{array}{c} \widetilde{\Delta}_0(k_i + k_j + k_k) \\ \| \\ \widetilde{\Delta}_0(k_\ell + k_m) \end{array} \right\} \left\{ \begin{array}{c} \widetilde{\Delta}_0(k_i + k_\ell + k_m) \\ \| \\ \widetilde{\Delta}_0(k_j + k_k) \end{array} \right\} \end{aligned} \quad (3.16)$$

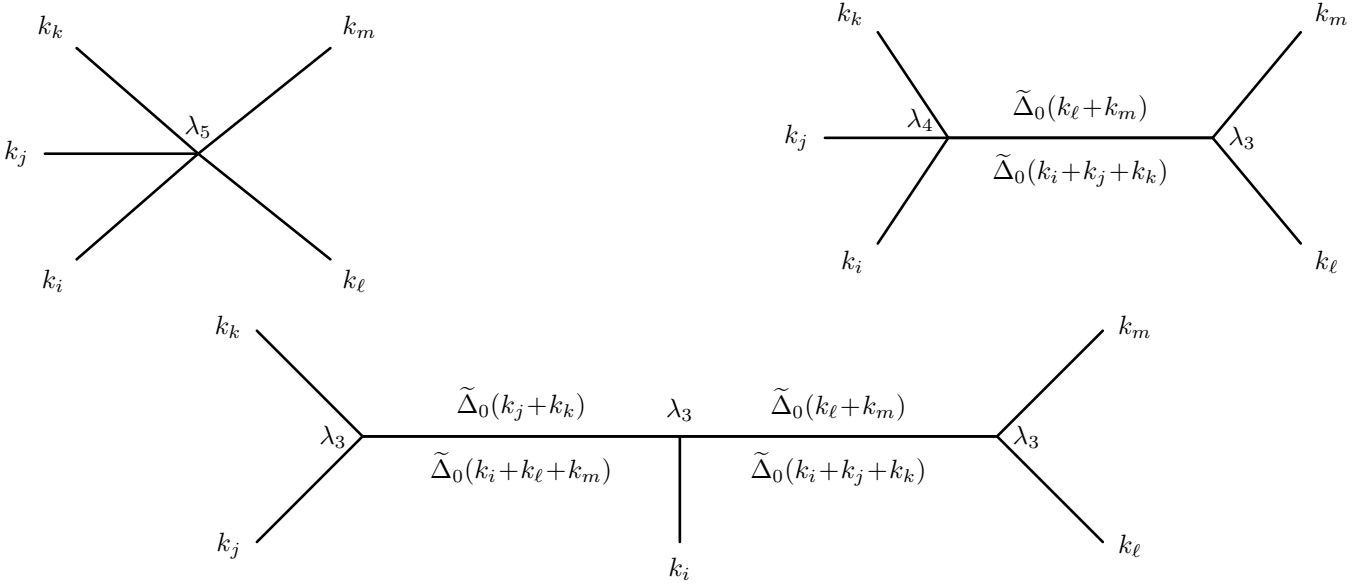

Figure 8: Connected 5-point tree diagrams.

$$\Rightarrow \quad -m^{-2}\mathcal{M}_5(k_1, k_2, k_3, k_4, k_5) \;=\; g_5 + g_3 g_4 \sum_{\substack{i,j,k \in \{1,\ldots,5\}}}^{i<j<k} \Delta_{ijk}$$
$$+ \frac{g_3^3}{2} \sum_{i=1}^{5} \sum_{\substack{j,k \notin \{i\}}}^{j<k} \sum_{\substack{\ell,m \notin \{i,j,k\}}}^{\ell<m} \left\{\begin{array}{c}\Delta_{ijk}\\ \| \\ \Delta_{\ell m}\end{array}\right\} \left\{\begin{array}{c}\Delta_{i\ell m}\\ \| \\ \Delta_{jk}\end{array}\right\},$$

(3.17)

cf. Fig. 8.

### 3.5.3 Poles

All propagators can be seen as 3-propagators, so all poles are of the form $a_i + a_j = 0$ (possibly after using on-shell conditions). Because of $S_5$-permutation symmetry, we may wlog. consider the pole $a_1 + a_2 = 0$. Momentum conservation then implies

$$\Rightarrow \quad (a_3, a_4, a_5) \;\propto\; (1, e^{\pm 2\pi i/3}, e^{\mp 2\pi i/3}).$$

(3.18)

### 3.5.4 Residue of $g_3 g_4$ diagram (3.17)

Then $(i,j) = (1,2)$ in eq. (3.17).

$$\begin{aligned}
\text{Res} \sum_{k \notin \{1,2\}} \Delta_{12k} &\stackrel{(3.7)}{=} \sum_{k \in \{3,4,5\}} \frac{a_1^2 a_k}{a_1^2 - a_k^2} \\
&\stackrel{(3.20)}{=} \frac{a_1^2}{D} \sum_{i,j,k \in \{3,4,5\}}^{\text{cycl.}} (a_1^2 - a_i^2)(a_1^2 - a_j^2) a_k \\
&\stackrel{(3.18)}{=} -\frac{a_1^4}{D} \sum_{i,j,k \in \{3,4,5\}}^{\text{cycl.}} (a_i^2 + a_j^2) a_k \\
&\stackrel{(3.18)}{=} \frac{a_1^4}{D} \sum_{k \in \{3,4,5\}} a_k^3 \\
&\stackrel{(3.18)}{=} 3\frac{a_1^4}{D} a_3 a_4 a_5,
\end{aligned} \quad (3.19)$$

where we have defined

$$D := \prod_{k=3}^{5} (a_1^2 - a_k^2). \quad (3.20)$$

### 3.5.5 Residue of $g_3^3$ diagram (3.17) where $i \notin \{1,2\}$

Then $(j,k) = (1,2)$ or $(\ell, m) = (1,2)$ in eq. (3.17). They give the same contribution, so assume the first option and multiply with 2:

$$\sum_{i \notin \{1,2\}} \sum_{\ell,m \notin \{i,1,2\}}^{\ell<m} \text{Res}\Delta_{i12}\Delta_{12} = -\text{Res}\sum_{i=3}^{5} \Delta_{i12} \stackrel{(3.19)}{=} -3\frac{a_1^4}{D} a_3 a_4 a_5, \quad (3.21)$$

which we already calculated in eq. (3.19).

### 3.5.6 Residue of $g_3^3$ diagram (3.17) where $i \in \{1,2\}$

Then $\{i,j\} = \{1,2\}$ or $\{i,\ell\} = \{1,2\}$ in eq. (3.17). They give the same contribution, so assume the first option and multiply with 2:

$$\begin{aligned}
\sum_{i=1}^{2} \sum_{k \notin \{1,2\}} \text{Res}\Delta_{12k} \sum_{\ell,m \notin \{1,2,k\}}^{\ell<m} \Delta_{i\ell m} &\stackrel{(3.5)\pm(3.7)}{=} \sum_{k=3}^{5} \frac{a_1^2 a_k}{a_1^2 - a_k^2} \sum_{\ell,m \notin \{1,2,k\}}^{\ell<m} \frac{a_\ell a_m}{a_\ell + a_m} \sum_{\pm} \frac{\pm a_1}{(a_\ell \pm a_1)(a_m \pm a_1)} \\
&\stackrel{(3.20)}{=} \frac{a_1^3}{D} \sum_{k=3}^{5} a_k \sum_{\ell,m \notin \{1,2,k\}}^{\ell<m} \frac{a_\ell a_m}{a_\ell + a_m} \underbrace{\sum_{\pm} \pm(a_\ell \mp a_1)(a_m \mp a_1)}_{= -2a_1(a_\ell + a_m)} \\
&= -2\frac{a_1^4}{D} \sum_{k=3}^{5} a_k \sum_{\ell,m \notin \{1,2,k\}}^{\ell<m} a_\ell a_m = -6\frac{a_1^4}{D} a_3 a_4 a_5.
\end{aligned} \quad (3.22)$$

### 3.5.7 Total residue of the 5-point function (3.17)

$$-m^{-2} \text{Res}\mathcal{M}_5(k_1, k_2, k_3, k_4, k_5) \stackrel{(3.19)+(3.21)+(3.22)}{=} \frac{a_1^4}{D} a_3 a_4 a_5 (3g_3 g_4 - 9g_3^3). \quad (3.23)$$

Eq. (3.23) yields that there are 2 possible ways to remove poles:

1. Case $g_3 = 0$.
2. Case $g_4 = 3g_3^2$.

We assume from now on that $\mathcal{M}_5$ has no poles and hence is constant. In fact we will assume that $g_5$ has been adjusted such that $\mathcal{M}_5 = 0$ has no particle production.

## 3.6 6-point amplitude

### 3.6.1 $\mathcal{M}_6 \sim \mathcal{M}_3 \Delta_0 \mathcal{M}_5$ factorization

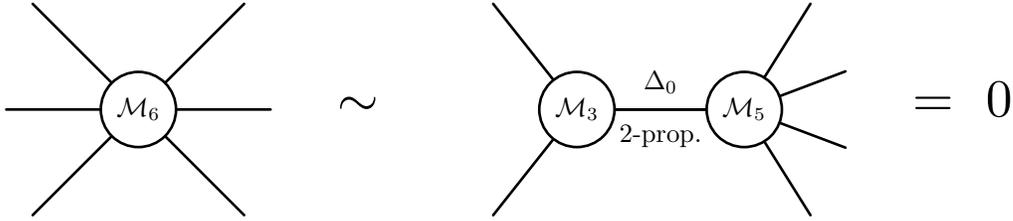

Figure 9: $\mathcal{M}_6 \sim \mathcal{M}_3 \Delta_0 \mathcal{M}_5$ factorization.

There are no poles associated with a 2-propagator since $\mathcal{M}_5 = 0$, cf. theorem 2.8 and Fig. 9.

### 3.6.2 No single-channel $\mathcal{M}_6 \sim \mathcal{M}_4 \Delta_0 \mathcal{M}_4$ factorization

A single pole clearly does not uniquely specify a factorization channel. Next consider e.g. a double pole $a_4 = -a_5 = a_6$ associated with a 3-propagator for the channel 456. Define

$$a_{456} := a_4 + a_5 + a_6 = -a_4. \tag{3.24}$$

Momentum conservation then implies

$$\Rightarrow \quad (a_1, a_2, a_3, a_{456}) \text{ are 2 pairs of opposite values } a_i + a_j = 0. \tag{3.25}$$

In other words: $\exists i \in \{1, 2, 3\}: \quad a_i = -a_4$. This is a simultaneous double pole for another channel, e.g. $i45$.

### 3.6.3 6-point diagrams

We cannot rely on factorization, so it is necessary to calculate the residues of the poles of $\mathcal{M}_6$. The only possible poles are associated with a 3-propagator (3.5), i.e. of the form $a_i + a_j = 0$. We therefore can exclude diagrams without 3-propagators, cf. Fig. 10. We only have to consider diagrams that contain 3-propagators, cf. Fig. 11. It is straightforward to check that the 6-point diagrams with 3-propagators group together according to their 3-propagator in an $\mathcal{M}_4 \Delta_0 \mathcal{M}_4$ block format. (Compare with Figs. 2 and 3.) In other words,

$$\mathcal{M}_6(k_1, k_2, k_3, k_4, k_5, k_6)|_{\text{3-prop}} = \frac{1}{2} \sum_{i,j,k \in \{1,\ldots,6\}}^{i<j<k} \mathcal{M}_4(k_i, k_j, k_k, \cdot)\, \widetilde{\Delta}_0(k_i+k_j+k_k) \sum_{\ell,m,n \notin \{i,j,k\}}^{\ell<m<n} \mathcal{M}_4(k_\ell, k_m, k_n, \cdot). \tag{3.26}$$

$$\Rightarrow \quad -m^{-2}\mathcal{M}_6(k_1, k_2, k_3, k_4, k_5, k_6)|_{\text{3-prop}} = \frac{1}{2} \sum_{i,j,k \in \{1,\ldots,6\}}^{i<j<k} \mathcal{M}_4(k_i, k_j, k_k, \cdot)\, \Delta_{ijk} \sum_{\ell,m,n \notin \{i,j,k\}}^{\ell<m<n} \mathcal{M}_4(k_\ell, k_m, k_n, \cdot). \tag{3.27}$$

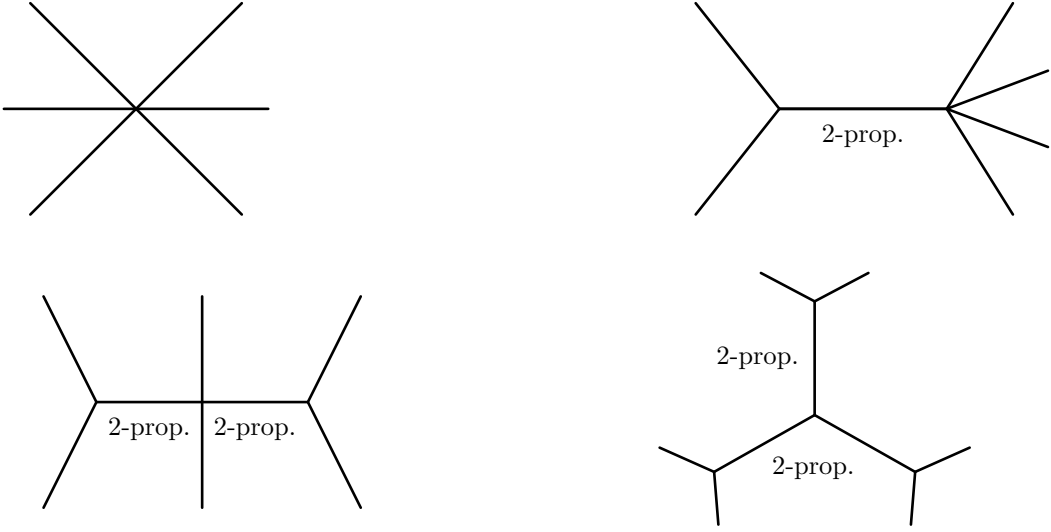

Figure 10: 6-point diagrams with no 3-propagator.

### 3.6.4 Poles

Because of $S_6$-permutation symmetry, we may wlog. consider the pole $a_1 + a_2 = 0$. Momentum conservation then implies

$$\Rightarrow \quad (a_3, a_4, a_5, a_6) \text{ are 2 pairs of opposite values } a_p + a_q = 0. \tag{3.28}$$

The dominant diagrams are, cf. Fig. 12,

$$-m^{-2}\mathcal{M}_6(k_1, k_2, k_3, k_4, k_5, k_6) \sim \sum_{k=3}^{6} \mathcal{M}_4(k_1, k_2, k_k, \cdot) \, \Delta_{12k} \sum_{\substack{\ell<m<n \\ \ell,m,n\notin\{1,2,k\}}} \mathcal{M}_4(k_\ell, k_m, k_n, \cdot) \tag{3.29}$$

$$\begin{aligned}
\Rightarrow \quad -m^{-2}\mathrm{Res}\mathcal{M}_6(k_1, k_2, k_3, k_4, k_5, k_6) &= \sum_{k=3}^{6} \mathcal{M}_4(k_1, k_2, k_k, \cdot)\frac{a_1^2 a_k}{a_1^2 - a_k^2} \sum_{\substack{\ell<m<n \\ \ell,m,n\notin\{1,2,k\}}} \mathcal{M}_4(k_\ell, k_m, k_n, \cdot) \\
&= \mathcal{M}_4(k_1, k_2, \cdot, \cdot) \sum_{k=3}^{6} \frac{a_1^2 a_k}{a_1^2 - a_k^2} \mathcal{M}_4(k_3, k_4, k_5, k_6) \\
&= 4 \text{ terms cancel in 2 opposite pairs } = 0.
\end{aligned} \tag{3.30}$$

Hence $\mathcal{M}_6$ has no poles and is constant. In fact we will assume that $g_6$ has been adjusted such that $\mathcal{M}_6 = 0$.

## 3.7 Higher-point amplitudes $\mathcal{M}_{n\geq 7} = 0$

The vanishing of higher-point amplitudes follows via induction in $n \geq 6$. Assume that $\mathcal{M}_{\leq n} = 0$. We want to prove that $\mathcal{M}_{n+1}$ cannot have any poles. The poles must come from internal propagators $\Delta_0$ via a single-channel factorization $\mathcal{M}_{n+1} \sim \mathcal{M}_r \Delta_0 \mathcal{M}_{n+3-r}$, where $r \in \{3, \ldots, n\}$, cf. theorem 2.8. It is easy to check that at least 1 of the 2 subamplitudes already vanishes.

## 3.8 Multi-Regge limit

The on-shell $n$-point amplitude $\mathcal{M}_n(k_1, \ldots, k_n)$ has in general $n-2$ independent lightcone-momenta $(a_1, \ldots, a_n)$ when we take momentum conservation into account.

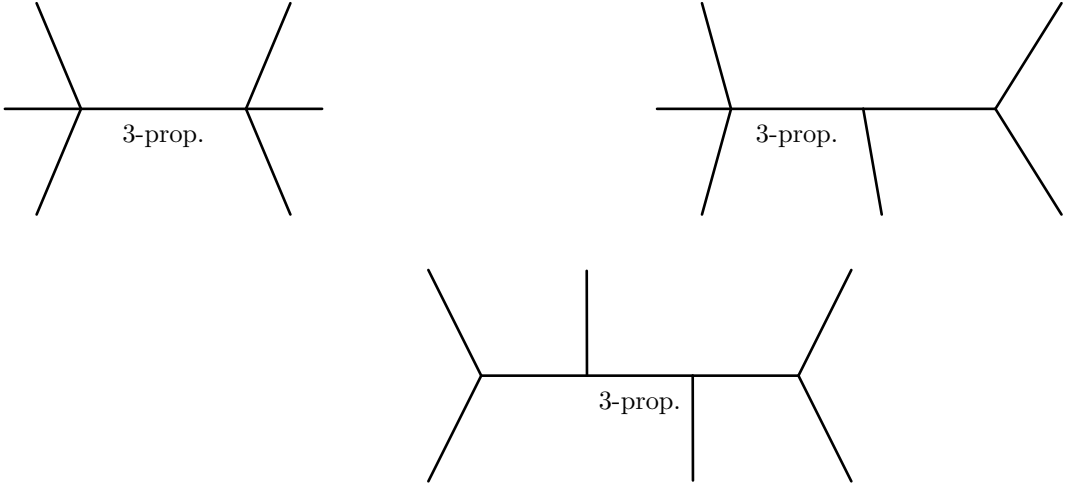

Figure 11: 6-point diagrams with a 3-propagator.

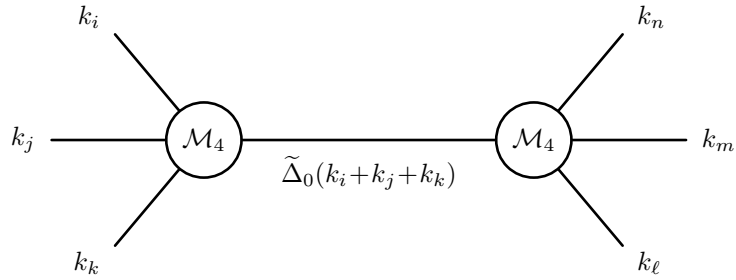

Figure 12: An $\mathcal{M}_4 \Delta_0 \mathcal{M}_4$ diagram.

**Definition 3.2** *The* **multi-Regge limit** *is a 1-parameter family of lightcone-momenta [5]*

$$a_i = x^{i-2}, \qquad i \in \{2, \ldots, n-1\}, \qquad n \geq 4, \tag{3.31}$$

where $x \gg 1$ is a free parameter.

The first and the last lightcone-momentum $a_1$ and $a_n$ are fixed (up to a 2-fold ambiguity) by the momentum conservation:

$$\begin{aligned} -a_1 - a_n &= \sum_{i=2}^{n-1} a_i \stackrel{(3.31)}{=} \sum_{i=2}^{n-1} x^{i-2} = x^{n-3}(1 + \mathcal{O}(x^{-1})), \\ -a_1^{-1} - a_n^{-1} &= \sum_{i=2}^{n-1} a_i^{-1} \stackrel{(3.31)}{=} \sum_{i=2}^{n-1} x^{2-i} = 1 + \mathcal{O}(x^{-1}), \end{aligned} \tag{3.32}$$

which leads to the same 2nd-order equation for $a_1$ and $a_n$. We fix the ambiguity of the 2 branches such that

$$a_1 = -1 + \mathcal{O}(x^{-1}), \qquad a_n = -x^{n-3}(1 + \mathcal{O}(x^{-1})). \tag{3.33}$$

### 3.8.1 Propagators

Consider an internal $r$-propagator in the multi-Regge limit

$$-\Delta_I := m^2 \widetilde{\Delta}_0(\sum_{i \in I} k_i)$$
$$= m^2 \widetilde{\Delta}_0(-\sum_{j \in J} k_j)$$
$$= \frac{1}{\sum_{i \in I} a_i \sum_{j \in J} a_j^{-1} + 1} \quad (3.34)$$
$$\stackrel{(3.31)}{=} \frac{1}{x^{i_r - 2}(1 + \mathcal{O}(x^{-1}))x^{2-j_1} + 1}$$
$$\rightarrow \begin{cases} 0 & \text{if } i_r > j_1 \\ 1 & \text{if } i_r < j_1 \end{cases} \quad \text{for} \quad x \to \infty,$$

where the index sets are

$$\begin{aligned} I &= \{i_1, i_2, \ldots, i_r\}, & 1 &= i_1 < i_2 < \ldots < i_r \leq n, \\ J &= \{1, \ldots, n\} \backslash I = \{j_1, j_2, \ldots, j_{n-r}\}, & 1 &< j_1 < j_2 < \ldots < j_{n-r} \leq n, \\ r &\in \{2, \ldots, n-1\}. \end{aligned} \quad (3.35)$$

Here the $n$ external lines corresponding to the index sets $I$ and $J$ are associated with opposite sides of the internal $r$-propagator.

We conclude the following statement.

**Statement 3.3** *In the multi-Regge limit the only non-vanishing internal $r$-propagator has ordered index sets*

$$I = \{1, 2, \ldots, r\} \quad \text{and} \quad J = \{r+1, r+2, \ldots, n\}. \quad (3.36)$$

### 3.8.2 Vertex expansion of a rooted tree

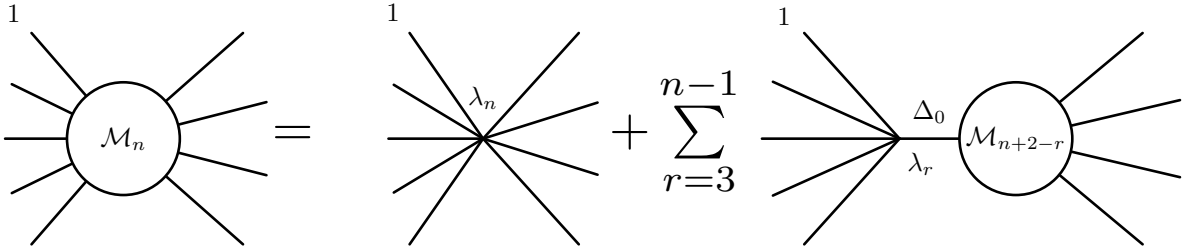

Figure 13: Vertex expansion of a rooted tree, cf. theorem 2.3. The root leg "1" is attached to a vertex, which in turn is attached to either i) external legs or ii) internal lines connected to tree amplitudes. Due to the multi-Regge limit, the vertex can at most be attached to 1 internal line, cf. statement 3.3. One should sum over possible types of vertices.

Now use a vertex expansion of a rooted tree [4, 5]

$$\mathcal{M}_n(k_1, \ldots, k_n) = -\lambda_n - \sum_{r=3}^{n-1} \lambda_r \underbrace{\widetilde{\Delta}_0(\sum_{i=1}^{r-1} k_i)}_{= m^{-2}} \mathcal{M}_{n+2-r}(\cdot, k_r, \ldots, k_n), \quad (3.37)$$

cf. Fig. 13. Next use that $\mathcal{M}_r = 0$ for $r \in \{5, 6, \ldots, n-1\}$ to obtain a 2-step recursion relation for $g_n$:

$$\Rightarrow \quad 0 = m^{-2}\mathcal{M}_n(k_1, \ldots, k_n) \stackrel{(3.37)}{=} -g_n - g_{n-1}m^{-2}\mathcal{M}_3(\cdot, k_{n-1}, k_n)$$
$$- g_{n-2}m^{-2}\mathcal{M}_4(\cdot, k_{n-2}, k_{n-1}, k_n) \quad (3.38)$$
$$\stackrel{(3.11)}{=} -g_n + g_{n-1}g_3 - g_{n-2}(-g_4 + g_3^2).$$

In the last line of eq. (3.38) we used that only 1 of the $s,t,u$ channels of $\mathcal{M}_4$ is consistent with the multi-Regge limit.

## 3.9 Result

The solution to the 2-step recursion relation (3.38) yields the following theorem 3.4.

**Theorem 3.4** *No particle production of a relativistic massive real scalar field theory (3.1) leads to 2 possible cases:*

1. *Case $g_3 = 0$:*

$$g_n = \frac{1 + (-1)^n}{2} g_4^{n/2-1} \quad \Leftrightarrow \quad \mathcal{V} = \frac{m^2}{g_4}\left[\cosh(\sqrt{g_4}\phi) - 1\right]. \quad (3.39)$$

   *This is the sinh-Gordon model if $g_4 > 0$ and the sine-Gordon model if $g_4 < 0$. It is the $a_1^{(1)}$ affine Toda model.*

2. *Case $g_4 = 3g_3^2$:*

$$g_n = \frac{2^n + 2(-1)^n}{6} g_3^{n-2} \quad \Leftrightarrow \quad \mathcal{V} = \frac{m^2}{6g_3^2}\left[\exp(2g_3\phi) + 2\exp(-g_3\phi) - 3\right]. \quad (3.40)$$

   *This is the Bullough-Dodd model, which is the $a_2^{(2)}$ affine Toda model.*

The above list is precisely rank-1 affine Toda field theories that satisfy assumption 2.2 [9].

## Acknowledgement

The author thanks Michal Pazderka and Linus Wulff for fruitful discussions. The work of K.B. is supported by the Czech Science Foundation (GACR) under the grant no. GA20-04800S for Integrable Deformations.